\documentclass[conference]{IEEEtran}
\IEEEoverridecommandlockouts

\usepackage{cite}
\usepackage{amsmath,amssymb,amsfonts}
\usepackage{algorithmic}
\usepackage{graphicx}
\usepackage{textcomp}
\usepackage{xcolor}
\usepackage{multirow}
\usepackage{float}
\usepackage[printonlyused,withpage]{acronym}
\usepackage{siunitx}
\usepackage{booktabs}
\usepackage{xcolor}
\acrodef{HGR}{Hand gesture recognition}
\acrodef{NNs}{neural networks}
\acrodef{SNNs}{spiking neural networks}
\acrodef{ANNs}{artificial neural networks}
\acrodef{FMCW}{frequency-modulated continuous-wave}
\acrodef{PLL}{phase-locked loop}
\acrodef{VCO}{voltage-controlled oscillator}
\acrodef{IF}{intermediate frequency}
\acrodef{ADC}{analog-to-digital converter}
\acrodef{RAF}{Resonate-and-fire}
\acrodef{FFTs}{fast Fourier transforms}
\acrodef{DFT}{discrete Fourier transform}
\acrodef{RMS}{Root Mean Square}
\acrodef{GRU}{gated recurrent unit}
\acrodef{RNN}{recurrent neural network}
\def\BibTeX{{\rm B\kern-.05em{\sc i\kern-.025em b}\kern-.08em
		T\kern-.1667em\lower.7ex\hbox{E}\kern-.125emX}}

\makeatletter
\newcommand{\linebreakand}{%
\end{@IEEEauthorhalign}
\hfill\mbox{}\par
\mbox{}\hfill\begin{@IEEEauthorhalign}
}
\makeatother
\begin{document}

\title{Resonate-and-Fire Spiking Neurons for Target Detection and Hand Gesture Recognition: A Hybrid Approach\\
\thanks{This work has received funding from the German Federal Ministry of Education and Research under the funding code 16MEE0111k and from the ECSEL Joint Undertaking (JU) under grant agreement No 876925. The JU receives support from the European Union’s Horizon 2020 research and innovation program and France, Belgium, Germany, Netherlands, Portugal, Spain, and Switzerland.}
}
\author{\IEEEauthorblockN{Ahmed Shaaban}
	\IEEEauthorblockA{\textit{University of Erlangen-Nuremberg} \\
		\textit{Infineon Technologies AG}\\
		Munich, Germany \\
		ahmed.shaaban@fau.de}
	\and
	\IEEEauthorblockN{Zeineb Chaabouni}
	\IEEEauthorblockA{
		\textit{Infineon Technologies AG}\\
		Munich, Germany \\}
	\and
	\IEEEauthorblockN{Maximilian Strobel}
	\IEEEauthorblockA{\textit{Infineon Technologies AG} \\
		Munich, Germany \\}
	\and
	\IEEEauthorblockN{Wolfgang Furtner}
	\IEEEauthorblockA{\textit{Infineon Technologies AG} \\
		Munich, Germany \\}
	\linebreakand
	\IEEEauthorblockN{Robert Weigel}
	\IEEEauthorblockA{\textit{Institute for Electronics Engineering} \\
		\textit{University of Erlangen-Nuremberg}\\
		Erlangen, Germany \\}
	\and
	\IEEEauthorblockN{Fabian Lurz}
	\IEEEauthorblockA{\textit{Chair of Integrated Electronic Systems} \\
		\textit{Otto-von-Guericke-University Magdeburg}\\
		Magdeburg, Germany \\
	}
}
\maketitle
\begin{abstract}
Hand gesture recognition using radar often relies on computationally expensive fast Fourier transforms. This paper proposes an alternative approach that bypasses fast Fourier transforms using resonate-and-fire neurons. These neurons directly detect the hand in the time-domain signal, eliminating the need for fast Fourier transforms to retrieve range information. Following detection, a simple Goertzel algorithm is employed to extract five key features, eliminating the need for a second fast Fourier transform. These features are then fed into a recurrent neural network, achieving an accuracy of 98.21\% for classifying five gestures. The proposed approach demonstrates competitive performance with reduced complexity compared to traditional methods.
\end{abstract}
\begin{IEEEkeywords}
Resonate-and-Fire, Spiking Neural Networks, Gesture Recognition, FMCW Radar, Machine Learning
\end{IEEEkeywords}
\section{Introduction}
\ac{HGR} is a captivating alternative to traditional click-and-touch interfaces~\cite{Radar1}, particularly for wireless control in diverse applications, including smart TVs, automotive systems, and intelligent audio devices. Radar-based \ac{HGR} solutions have gained significant traction due to their inherent privacy advantages, operational robustness in various illumination and environmental conditions, cost-effectiveness, reduced energy consumption, and seamless integration into end-user products. In contrast, camera-based \ac{HGR} systems face limitations regarding privacy concerns, higher costs, and larger physical footprints [2]. To address these limitations, several \ac{ANNs}-based solutions for radar \ac{HGR} have been proposed \cite{Radar3, Radar4, Radar5, Max}. However, most of these approaches rely on conventional radar signal processing pipelines, which require computationally expensive two-dimensional \ac{FFTs} to extract range (radial distance) and Doppler (radial velocity) information.

Time-domain-based \ac{HGR} solutions have emerged to circumvent the computational burden of \ac{FFTs}. These approaches feed the time-domain radar data into \ac{NNs}, enabling implicit feature extraction and classification within the network architecture. Additionally, they utilize \ac{SNNs} due to their sparse and computationally efficient nature, further contributing to improved computational efficiency \cite{SNNs1, SNNs2}.

\ac{RAF} neurons \cite{RAF} demonstrate great versatility across diverse domains. Research reveals their ability to encode radar signals into spikes in radar signal processing, enabling efficient interference detection \cite{Julian}. These same neurons excel at transforming audio signals into a frequency-selective representation, proving valuable for speech recognition tasks \cite{Auge}. Furthermore, recent studies explore the direct application of RAF neurons on time-domain radar data for hand target detection in \ac{HGR} systems \cite{RAF_1}.

Extending the work presented in \cite{RAF_1}, this paper proposes an approach for \ac{HGR} that leverages the unique capabilities of \ac{RAF} neurons. Our approach introduces three key contributions: We employ a layer of \ac{RAF} neurons directly on the time-domain radar data to detect the hand (target), bypassing conventional FFTs for radial distance information extraction. A computationally efficient feature extraction approach based on the Goertzel algorithm is utilized following hand detection. This allows extracting five informative features: radial distance, radial velocity, azimuth, elevation angles, and signal amplitude, eliminating the necessity for a second FFT typically used in conventional approaches. Finally, we implement a \ac{RNN} to classify five distinct hand gestures successfully.
To the best of our knowledge, this is the first work to employ \ac{RAF} neurons for developing a complete \ac{HGR} solution, demonstrating their potential for bypassing the \ac{FFTs} and achieving efficient gesture recognition.
\section{Radar System and Dataset}
\subsection{Radar System}\label{RadarSystem}
This work utilizes the BGT60TR13C \SI{60}{\giga\hertz} \ac{FMCW} radar from Infineon Technologies for gesture data collection~\cite{FMCW, FMCW_1}. The radar employs a single transmitting antenna and three receiving antennas arranged in an L-shape, enabling the calculation of azimuth and elevation angles using two receiver antennas.
The radar transmits chirps, which are linearly increasing frequency waveforms. These transmitted chirps reflect off targets and are received by the three receiving antennas.
The received signal is mixed with the transmitted signal, resulting in the \ac{IF} signal \cite{FMCW_2}. This \ac{IF} signal undergoes a two-step filtering process. A high-pass filter removes frequencies below \SI{100}{\kilo\hertz}, while a low-pass filter safeguards against aliasing with leakage of far-range targets by eliminating high-frequency interventions. This filtering ensures that the \ac{IF} signal occupies a significantly lower frequency range (\SI{100}{\kilo\hertz} - \SI{600}{\kilo\hertz}) than the transmitted signal. This reduction in frequency enables efficient digitization using an \ac{ADC} with a sampling rate of \SI{2}{\mega\hertz}. The digitized signals from the three receiving antennas form the final output from the radar.

The radar operates with a chirp configuration of 32 chirps per frame and a chirp repetition rate $T_c$ of \SI{0.3}{\milli\second}. Each chirp undergoes frequency modulation, sweeping from a start frequency $F_{\mathrm{min}}$ of \SI{58.5}{\giga\hertz} to a stop frequency $F_{\mathrm{max}}$ of \SI{62.5}{\giga\hertz}, resulting in a bandwidth $B$ of \SI{4}{\giga\hertz}. This configuration yields a range resolution of  $R_{\text{res}} = \frac{c_{\text{o}}}{2B}$, where ${c_{\text{o}}}$ is the speed of light, which translates to approximately \SI{0.038}{\meter}.
The chirps are sampled at a rate of \SI{2}{\mega\hertz} with 64 samples per chirp. This sampling scheme defines the maximum unambiguous range $R_{\text{max}}$ as $R_{\text{res}} \times \frac{N_s}{2}$, where ${N_s}$ is the number of samples, resulting in an $R_{\text{max}}$ of \SI{1.2}{\meter}.
Furthermore, the maximum measurable velocity is determined by $V_{\max} = \frac{c_0}{4\times F_c\times T_c}$, where $F_c$ represents the center frequency of the chirps. Thus, the maximum measurable velocity is approximately \SI{4.13}{m/s}.
Table~\ref{RadarParameters} summarizes the key operating parameters of the radar employed in this work.
\begin{table}[t!]
	\centering
	\caption{Radar Operating Parameters}
	\label{RadarParameters}
	\begin{tabular}{@{}cc@{} }
		\toprule
		\textbf{Parameter}                 & \textbf{Value} \\ \midrule
		Start frequency ($F_{\mathrm{min}}$)           & \SI{58.5}{\giga\hertz}      \\
		Stop frequency ($F_{\mathrm{max}}$)            & \SI{62.5}{\giga\hertz}       \\
		Center frequency ($F_{c}$)            & \SI{60.5}{\giga\hertz}       \\
		Bandwidth ($B$)                      & \SI{4}{\giga\hertz}          \\
		Number of samples per chirp ($N_{s}$)   & 64             \\
		Number of chirps per frame ($N_{c}$)    & 32             \\
		Number of frames per recording ($N_{f}$) & 100            \\
		Chirp repetition time ($T_{c}$)        & \SI{0.3}{\milli\second}         \\
		Frame repetition time ($T_{f}$)         & \SI{30}{\milli\second}         \\
		Sampling frequency ($F_{s}$)         & \SI{2}{\mega\hertz}         		\\  \bottomrule
	\end{tabular}
\end{table}

\subsection{Dataset}\label{Dataset}
This work utilizes a gesture dataset recorded by eight individuals (six males and two females) aged 18-29 years. The recordings were conducted with minimal supervision in diverse environments, including silence boxes, libraries, gyms, regular rooms, kitchens, and office spaces, encompassing various static objects like walls, chairs, and tables. The dataset comprises a total of 18,299 recordings containing the following gestures: SwipeLeft (hand movement from right to left), SwipeRight (hand movement from left to right), SwipeUp (hand movement from down to up), SwipeDown (hand movement from up to down), and Push (hand movement towards the radar followed by a slight retraction).

A single gesture recording is comprised of 100 frames. Each frame contains 32 individual chirps, and each chirp consists of 64 samples. As detailed in Table~\ref{RadarParameters} and Section~\ref{RadarSystem}, the chirp repetition time is 0.3 ms, and the frame repetition time is 30 ms. Consequently, each gesture recording takes approximately 3 seconds. Considering the three receiving antennas, the final data shape is (18,299 recordings, 3 antennas, 100 frames, 32 chirps, 64 samples). An extended version of the dataset is available at \cite{Sarah}. 
\section{Proposed Recognition System}
\subsection{Resonate-and-Fire Neurons: Frequency-Selective Processing without Training}
\ac{RAF} neurons~\cite{RAF} offer an existing approach to modeling biological neuron dynamics by employing a damped oscillation to represent the neuron state variable (membrane potential). This characteristic enables them to function as frequency filters. Drawing inspiration from harmonic oscillators, \ac{RAF} neurons incorporate a spiking mechanism, allowing them to directly convert continuous-time radar signals into frequency-dependent spike representations. This eliminates the need for complex feature extraction steps such as a FFT, streamlining the processing pipeline.
The core principle of \ac{RAF} neurons lies in their inherent resonance frequency. When an input signal aligns in frequency with the neuron's resonance, and its magnitude surpasses a specific threshold with the correct phase, the neuron generates spikes, signifying resonance detection. In particular, \ac{RAF} neurons exhibit a temporal aspect, incorporating the history of input signals into their decision-making process, as their state variable component relies on previous time step values, ensuring oscillation and spiking at the resonant frequency. Additionally, the neuron's internal parameters, such as the decay rate $\alpha$, influence its behavior by controlling the speed of returning to its resting state after firing. This parameter, along with the spiking threshold $V_{th}$, is tuned to achieve the desired resonance characteristics. These parameters ensure non-spiking activity at non-resonant frequencies and robustness against noise in the input signals.
In essence, \ac{RAF} neurons act as powerful frequency-selective filters, transforming time-domain radar signals directly into spike representations based on the resonance between the input and the neuron's internal properties. This capability is also achieved without explicit training, making \ac{RAF} neurons valuable for various signal-processing applications.

\subsection{Refined Gesture Labeling and Efficient Target Detection using RAF Neurons}\label{TargetDetection}
Building upon Section~\ref{Dataset}, while a gesture recording encompasses 100 frames (approximately 3 seconds), only a subset, lasting hundreds of milliseconds, captures the actual hand gesture execution. Traditionally, labeling all frames within a recording for gesture classification leads to unrealistic solutions with increased latency and computational burden. 

This work utilizes an innovative approach that leverages frame-wise labeling \cite{Max, RAF_1}. The system facilitates real-time predictions per frame by identifying and labeling only the frames containing genuine hand gestures (”Gesture Frames”) and differentiating them from accompanying Background frames, which is crucial for real-world scenarios. The network prioritizes feature extraction from these gesture frames during NN training for accurate classification. Consequently, this method enables real-time gesture detection upon user approach and gesture execution while simultaneously classifying the Background state during inactivity.

To achieve this, this work presents a computationally efficient advancement upon prior work~\cite{RAF_1}. While the previous approach employed 32 \ac{RAF} neurons~\cite{Lava} to analyze the 64 samples from 10 chirps within each frame, we enhanced this process by utilizing the same neurons to explore the 64 samples from a subset of only 3 chirps within each frame. These neurons identify the resonance frequency exhibiting the most prominent spike while closest to the radar. This approach capitalizes on the observation that gesture frames often contain two dominant frequencies: one for the hand (closer to the radar) and one for the body. The system effectively identifies the target (hand) where a gesture is executed by focusing on the fundamental resonance frequency closest to the radar. Remarkably, this method reduces computational demands by bypassing traditional range-\ac{FFTs} typically applied to the 64 samples from 32 chirps within a frame.

Additionally, by applying this processing to each frame within a recording,  the frame ($Gesture_{Frame}$) with the earliest occurrence of the hand's fundamental resonance frequency can be identified, indicating the precise moment of gesture execution while being closest to the radar. This frame typically represents the midpoint for Swipe gestures, while Push gestures signify the near end of the movement.

In essence, the approach of using 32 \ac{RAF} neurons enables frame-level target (hand) detection and refined gesture labeling directly from time-domain radar data while still bypassing traditional range-\ac{FFTs}.
\subsection{Goertzel Algorithm}\label{Goertzel}
As mentioned in Section~\ref{TargetDetection}, 32 \ac{RAF} neurons are used to identify the hand's resonance frequency (target) within each gesture frame. This information is then utilized to perform frequency analysis using the Goertzel Algorithm \cite{Goertzel, Goertzel_1}, focusing solely on the hand's specific resonance frequency. This approach eliminates the need to analyze irrelevant frequencies in the gesture frame.
Recall that a single gesture frame comprises data from 3 antennas, 32 chirps, and 64 samples. The Goertzel Algorithm analyzes only the 64 samples at the hand's resonance frequency for each antenna and chirp. This algorithm functions by:
\begin{itemize}
	\item Initializing a complex coefficient: This coefficient corresponds to the hand's resonance frequency obtained from the RAF neuron analysis.
	\item Sample-by-sample processing: The algorithm iterates through each sample within the 64-sample window, applying a second-order recursive filter.
	\item Output generation: Upon processing all samples, the algorithm generates a complex number analogous to the outcome of evaluating the \ac{DFT} at just the hand's resonance frequency.
\end{itemize}
Therefore, the Goertzel Algorithm provides a set of complex numbers representing the \ac{DFT} coefficients in the dimension of (3 antennas, 32 chirps), effectively capturing the hand's signature within each gesture frame.
\begin{figure*}[t!]
	\centering
	\footnotesize
	\def\svgwidth{\textwidth}
	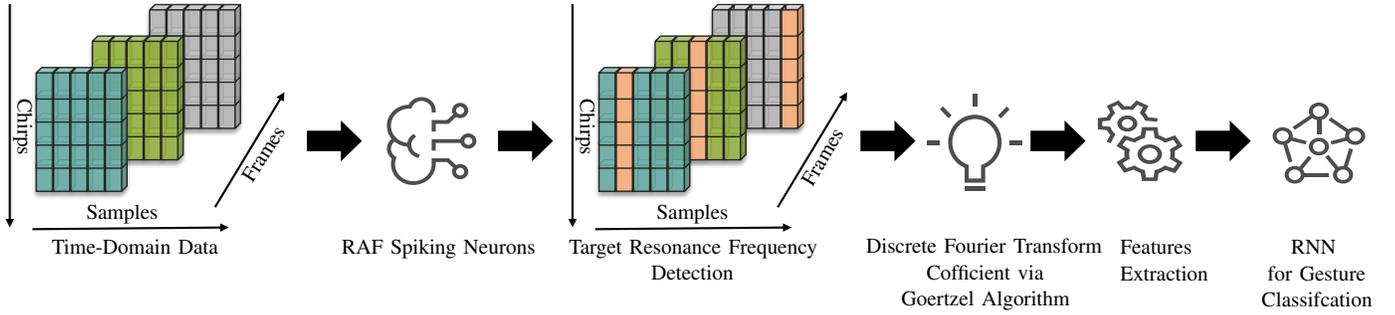
	\caption{Hand gesture recognition approach:  The approach utilizes a layer of 32 resonate-and-fire (RAF) neurons to effectively detect the hand's resonance frequency. Subsequently, the Goertzel algorithm is employed at this specific frequency to calculate the discrete Fourier transform coefficient. This coefficient, along with simple angular and phase estimations, allows for extracting five essential features. These features are then fed into a recurrent neural network, enabling the system to classify between five distinct hand gestures.}
	\label{fig:ProposedAlgo}
	\vspace{-\baselineskip}%
\end{figure*}
\subsection{Features Extraction}\label{Features}
As a next step, to extract informative features enriching each frame with intuitive gesture-specific characteristics, the following processing steps are employed on each time-domain radar frame:
\begin{enumerate}
	\item Preprocessing: The signal undergoes mean removal along both the sample (fast-time) and chirp (slow-time) dimensions to eliminate DC components and reflections from stationary objects~\cite{MTI}. Additionally, min-max normalization is applied to each frame, scaling all values between 0 and 1 to facilitate subsequent feature extraction.
	\item Target Detection and Feature Extraction:
	\begin{itemize}
		\item 32 \ac{RAF} Neurons: This layer identifies the hand's resonance frequency (target) within each frame and the corresponding spiking neuron index (providing range information).
		\item Goertzel Algorithm: Applied solely at the hand's resonance frequency, this algorithm efficiently extracts the \ac{DFT} coefficients for all three antennas and 32 chirps.
		\item Doppler Information: Utilizing the Goertzel output, phase differences between the first and second chirps are calculated and averaged across antennas to enhance extracted information.
		\item Azimuth Angle: The phase monopulse algorithm \cite{Monopulse} determines the azimuth angle. This algorithm leverages the phase difference between the third and first antenna, averaged across chirps, for improved accuracy.
		\item Elevation Angle: Similar to the azimuth angle calculation, the phase difference between the third and second antenna is used, averaged across chirps, to determine the elevation angle.
		\item \ac{RMS} Amplitude: This value is computed across all antennas and chirps using the Goertzel algorithm output.
	\end{itemize}
\end{enumerate}
Consequently, this comprehensive approach yields a set of five informative features per frame, all calculated at the hand's location (target bin): 1. Spiking Neuron Index: Corresponding to the hand's resonance frequency (radial distance). 2. Doppler Information: Reflecting hand motion characteristics (radial velocity). 3. Azimuth Angle: Indicating the hand's horizontal position. 4. Elevation Angle: Indicating the hand's vertical position. 5. \ac{RMS} Amplitude: Representing the hand's signal strength.
This combination of features effectively captures crucial information for robust gesture classification. Fig.~\ref{fig:ProposedAlgo} presents the workflow of the proposed HGR system.
\subsection{Architecture}
Capitalizing on the feature extraction process detailed in Section~\ref{Features}, a final feature representation with a shape of (100, 5) for each gesture recording is obtained. This arises from combining 100 temporal frames per recording and extracting five features per frame.
Recognizing the inherent temporal nature of the data, a \ac{GRU} is employed as the model of choice. As a variant of \ac{RNN}, \ac{GRU}s excel at capturing temporal dependencies within sequential data. Moreover, the \ac{GRU} architecture will be able to focus on learning from features corresponding to actual hand movement frames, effectively distinguishing them from Background labeled frames. The employed \ac{GRU} network remains computationally efficient, featuring a hidden state with an output size of 16 and a total of 1206 trainable parameters.
\section{Experimental Setup}
\subsection{Dataset Preparation}
The dataset described in Section~\ref{Dataset}, encompassing 18,299 gesture recordings, is split into training (10,678 recordings), validation (3,022 recordings), and testing (4,599 recordings) sets.
Following the feature extraction process outlined in Section~\ref{Features}, the extracted five-dimensional features undergo normalization to enhance the training efficiency and convergence of the \ac{GRU} network. Remarkably, these features exhibit varying value ranges. To address this, a StandardScaler is employed to normalize the features, effectively scaling them to have a zero mean and unit variance. This crucial step promotes efficient training by ensuring all features contribute equally to the learning process and prevents specific features with larger scales from dominating the network's learning.
This normalization is performed exclusively on the training data. Subsequently, the resulting normalization parameters transform features from all three datasets (training, validation, and testing) before they are presented to the \ac{GRU} network. This approach ensures consistent feature representation across all datasets while leveraging the knowledge gained from the training data.

Building upon the initial approach detailed in~\cite{RAF_1} and further elaborated in Section~\ref{TargetDetection}, this work implements a refined labeling strategy for gesture recordings. This strategy relies on the 32 \ac{RAF} neurons to identify the $Gesture_{Frame}$, where a hand gesture is performed at its closest proximity to the radar.
Following the identification of the $Gesture_{Frame}$, neighboring frames are meticulously labeled to capture the temporal dynamics of the hand movement associated with each gesture type. For Swipe gestures, 4 frames before and 4 after the $Gesture_{Frame}$ are labeled as containing the gesture execution. For the Push gestures, 5 frames before and 3 after the $Gesture_{Frame}$ are labeled as containing the gesture execution. All remaining frames within a gesture recording are classified as belonging to the Background class.

\subsection{Network Training}\label{Training}
Prior to network training, a hyperparameter optimization process is performed to identify the optimal values for the \ac{RAF} neurons' decay rate $\alpha$ and threshold $V_{th}$. This optimization process results in selecting 0.018 and 0.02 for the decay rate and threshold, respectively.
The Adam optimizer is employed for network optimization, utilizing a learning rate of $1.58 \times 10^{-3}$ and a weight decay of $1.6 \times 10^{-5}$~\cite{Adam}. The network's output is processed through a LogSoftMax function to enhance classification performance. This is followed by applying the Negative LogLikelihood Loss function to calculate the loss during training.
Early stopping is implemented to stand against overfitting~\cite{EarlyStopping}. Early stopping monitors the validation loss during training and terminates the training process if there is no improvement in the validation loss for a predefined patience period (10 epochs in this case). This safeguard prevents the model from memorizing the training data and ensures better generalization to unseen data.
The networks have been trained for 150 epochs, using a batch size of 32, and the entire training process is repeated ten times to account for potential randomness inherent in NN optimization. Each repetition utilizes a distinct random seed, effectively mitigating the impact of initialization variations. Further, the training process identifies and retains the "best" model, achieving the lowest validation loss for subsequent evaluation and analysis. Table~\ref{HyperparametersOverview} outlines the hyperparameters utilized for the \ac{RAF} neurons and the \ac{GRU} model training.
\begin{table}[t!]
	\centering
	\caption{Hyperparameter Settings}
	\label{HyperparametersOverview}
	\begin{tabular}{cc}
		\toprule
		\multicolumn{2}{c}{\textbf{General Hyperparameters}} \\
		\midrule
		Batch size               & 32                   \\
		Epochs                   & 150                  \\
		Adam learning rate       & $1.58\times 10^{-3}$  \\
		Adam weight decay        & $1.6\times 10^{-5}$ \\
		\midrule
		\multicolumn{2}{c}{\textbf{RAF Specific Hyperparameters}} \\
		\midrule
		Beta ($\alpha$)                    & 0.018                  \\
		Threshold ($V_{th}$)   & 0.02                 \\
		\bottomrule
	\end{tabular}
\end{table}
\subsection{Testing and Evaluation}\label{Testing}
Following the training process, the ten "best models" identified based on the lowest validation loss are used for evaluation. These models are loaded and tested on the previously unseen testing dataset, ensuring a robust assessment of the framework's generalization capabilities. The reported results reflect the average performance across all ten models, comprehensively representing the overall system's effectiveness.
\section{Results and Discussion}
\subsection{Evaluation of the Resonant-and-Fire-based Gesture Recognition System}
Following the implementation of the refined labeling strategy, the \ac{GRU} network undergoes training as outlined in Section~\ref{Training} and evaluation as described in Section~\ref{Testing}. These processes utilize features extracted as detailed in Section~\ref{Features} and split according to the methodology outlined in Section~\ref{Dataset}. The ensemble of the ten best \ac{GRU} models achieves a mean test accuracy of 98.21\%, demonstrating the effectiveness of the proposed approach.
\subsection{Further Analysis}\label{Further Analysis}
To evaluate the effectiveness of our proposed system, particularly the capability of \ac{RAF} neurons to detect hand resonance frequencies, a comparison is conducted with the state-of-the-art target detection approach from~\cite{Max}. This approach employs range-\ac{FFTs}, filtering, and thresholding to isolate the hand from surrounding objects. The identified hand range bin is then utilized for feature extraction using the Goertzel algorithm (as detailed in Sections~\ref{Goertzel} and~\ref{Features}).
Furthermore, the \ac{GRU} network architecture undergoes training and evaluation following the methodology outlined in Sections~\ref{Training} and~\ref{Testing}, ensuring a fair comparison. The ten best \ac{GRU} models in this comparative analysis achieve an accuracy of 98.11\%.\newline

To further explore the impact of feature extraction on system performance, an additional set of features inspired by~\cite{Max} is evaluated. This investigation utilizes the hand range bin identified using the conventional FFT-based approach and extracts the following features:
\begin{itemize}
	\item Doppler information: Obtained by applying a second FFT to the range information of the hand bin.
	\item Azimuth and Elevation Angles: Estimated from the Doppler information using a phase monopulse, similar to the proposed approach (Section~\ref{Features}).
	\item Signal Amplitude: Calculated as the peak value of the Doppler information at the hand bin.
\end{itemize}
These five features are then fed into the \ac{GRU} network, which undergoes training and evaluation following the same methodology as the proposed approach.
This investigation reveals that the conventional approach achieves an accuracy of 98.10\%.

Table~\ref{Comparison} summarizes the performance of the proposed gesture recognition system in comparison with the two alternative approaches: (1) the FFT-based hand (target detection) + Goertzel feature extraction and (2) the conventional FFT-based gesture recognition system.
\begin{table}[t!]
	\centering
	\renewcommand{\arraystretch}{1.2} 
	\caption{Comparison of Different HGR Approaches}
	\label{Comparison}
	\begin{tabular}{@{}p{6cm}|c@{}}
		\toprule
		\textbf{HGR Approach} & \textbf{Mean Test Accuracy} \\ \midrule
		Proposed Approach & \textbf{98.21} \% \\ \cline{2-2}
		Target detection using FFT + Goertzel & \textbf{98.11} \% \\ \cline{2-2}
		Target detection and feature extraction using FFT & \textbf{98.10} \% \\ \bottomrule
	\end{tabular}
\end{table}
\subsection{Discussion}
The proposed \ac{RAF}-based gesture recognition system achieves a mean test accuracy of 98.21\%, demonstrating its effectiveness in accurately classifying five distinct hand gestures. This high performance can be attributed to several key strengths:

\textbf{Efficient Hand Detection}: Utilizing \ac{RAF} neurons eliminates the need for complex conventional hand detection techniques like range-FFTs, filtering, and thresholding. This simplifies the processing pipeline while maintaining comparable performance, as evidenced by the comparison with the traditional method, achieving 98.11\% accuracy. This finding validates the effectiveness of \ac{RAF} neurons as an alternative for hand detection in gesture recognition systems.

\textbf{Simplified Feature Extraction}: The proposed approach uses the computationally cheap Goertzel algorithm to exploit the resonance frequencies extracted by \ac{RAF} neurons, potentially leading to more meaningful features than conventional methods.

\textbf{Competitive Performance with Reduced Complexity}: Section~\ref{Further Analysis} demonstrates the proposed system's competitive performance despite its streamlined processing pipeline. While achieving a comparable accuracy of 98.21\%, it avoids the complexities (2D \ac{FFTs}) of conventional hand detection and feature extraction techniques employed by the traditional approach (98.10\% accuracy). The proposed approach presents an effective alternative to the methods outlined in \cite{Max} by reducing computational complexity. In conventional hand detection, applying \ac{FFTs} on the entire samples dimension of each chirp incurs a complexity of $N_{c} * N_{s} * log_{2}(N_{s})$, where $N_{c}$ represents the number of chirps (32) and $N_{s}$ represents the number of samples per chirp (64). This approach, however, achieves a lower complexity of $N_{RAF} * N_{S-RAF}$. Here, $N_{RAF}$ denotes the number of RAF neurons (32), and $N_{S-RAF}$ signifies the number of samples utilized within the \ac{RAF} solution (192, drawn from 3 chirps). Furthermore, the proposed approach applies the Goertzel algorithm for each chirp at the hand resonance frequency, estimating amplitude and phase. Doppler information is then calculated as the phase difference between the first two chirps, incurring a complexity of $O(N_{c})$ (where $N_{c}$ remains 32) – an improvement compared to the conventional Doppler-FFT applied at the range information of the hand range bin, which exhibits a complexity of $N_{c} * log_{2}(N_{c})$.

This simplification translates to the noteworthy potential for reduced computational cost and implementation complexity when deployed on neuromorphic hardware. Neuromorphic platforms can directly exploit the inherent spiking nature of \ac{RAF} neurons, leading to a more efficient processing pipeline. This advantage is crucial for real-world applications with resource constraints and low latency requirements.

\section{Conclusion}
This work proposed a streamlined hand gesture recognition system that bypasses complex fast Fourier transform processing. Utilizing untrained resonate-and-fire (RAF) neurons for efficient hand detection and a simple Goertzel algorithm for feature extraction, it achieves competitive accuracy of 98.21\% for five gestures. This matches the performance of conventional approaches, showcasing the effectiveness of RAF neurons and the proposed approach. Particularly noteworthy is the potential for implementation on neuromorphic hardware. The spiking nature of RAF neurons offers significant promise for reduced computational costs. The future exploration of a fully spiking system employing the proposed simplified processing pipeline and a spiking neural network opens exciting avenues.

\end{document}